\documentclass[conference]{IEEEtran}
\usepackage[utf8]{inputenc}
\usepackage[T1]{fontenc}
\usepackage[english]{babel}
\usepackage{amsmath,amssymb}
\usepackage{booktabs}
\usepackage{graphicx}
\usepackage{xcolor}
\usepackage{pgfplots}
\pgfplotsset{compat=1.18}
\usepgfplotslibrary{statistics}
\usepackage{stfloats}
\usepackage{float}

\usepackage{tikz}
\usetikzlibrary{arrows.meta,positioning,calc}
\usepackage[hidelinks]{hyperref}
\usepackage{url}

\usepackage[normalem]{ulem}
\newcommand{\areqOWE}{223}     %
\newcommand{\arespOWE}{206}    %
\newcommand{\areqPQC}{1419}    %
\newcommand{\arespPQC}{1306}   %
\newcommand{\areqFiveOneTwo}{1033}\newcommand{\arespFiveOneTwo}{984}\newcommand{\ovhFiveOneTwo}{1588} %
\newcommand{\areqTenTwoFour}{1807}\newcommand{\arespTenTwoFour}{1790}\newcommand{\ovhTenTwoFour}{3168} %
\newcommand{\nfallback}{1}\newcommand{\nfallbackTot}{254} %
\newcommand{\dtassocOWE}{1.17} %
\newcommand{\dtassocPQC}{1.41} %
\newcommand{\dtassocDelta}{0.23} %
\newcommand{\dtassocPct}{20}     %
\newcommand{\mwupval}{1.8\times10^{-7}} %
\newcommand{\nlat}{30}           %
\newcommand{\bpOWElw}{0.998}\newcommand{\bpOWEq}{1.056}\newcommand{\bpOWEQ}{1.281}\newcommand{\bpOWEuw}{1.433}
\newcommand{\bpPQlw}{1.193}\newcommand{\bpPQq}{1.323}\newcommand{\bpPQQ}{1.575}\newcommand{\bpPQuw}{1.909}
\newcommand{\bpPQout}{2.178} %
\newcommand{\wcOWE}{2.6}
\newcommand{\wcPQC}{2.9}
\newcommand{\tkgFiveOneTwo}{44.5}\newcommand{\tenFiveOneTwo}{46.3}\newcommand{\tdeFiveOneTwo}{45.5}\newcommand{\thkFiveOneTwo}{3.8}\newcommand{\ttotFiveOneTwo}{140}
\newcommand{\tkgSevenSix}{62.8}\newcommand{\tenSevenSix}{63.9}\newcommand{\tdeSevenSix}{57.4}\newcommand{\thkSevenSix}{3.8}\newcommand{\ttotSevenSix}{188}
\newcommand{\tkgTenTwoFour}{91.8}\newcommand{\tenTenTwoFour}{94.0}\newcommand{\tdeTenTwoFour}{75.5}\newcommand{\thkTenTwoFour}{4.1}\newcommand{\ttotTenTwoFour}{265}
\newcommand{\nmicro}{2000}
\newcommand{\envCPU}{AMD Ryzen 5 5500 (Zen 3, 3.6\,GHz base), 2 vCPUs in a VirtualBox VM}
\newcommand{\envRAM}{3.8\,GiB}
\newcommand{\envOS}{Kali GNU/Linux Rolling (kernel 6.12.13-amd64)}
\newcommand{\envOpenSSL}{OpenSSL 3.6.2}
\newcommand{\envCC}{GCC 14.2.0 (Debian 14.2.0-16), \texttt{-O2}}
\newcommand{\envHostapd}{hostapd/wpa\_supplicant 2.10 (patched)}
\newcommand{\envDriver}{mac80211\_hwsim}

\begin{document}

\title{Post-Quantum Migration of Wi-Fi Key Establishment: A Hybrid ML-KEM Design for OWE and FILS, Validated on OWE with \textit{hostapd}}

\author{
\IEEEauthorblockN{Jes\'us R. Abascal}
\IEEEauthorblockA{Universidad de Cantabria\\ Avenida de los Castros s/n\\ Santander, Spain\\ jesus.abascal@alumnos.unican.es}
\and
\IEEEauthorblockN{Ana I. G\'omez}
\IEEEauthorblockA{Universidad Rey Juan Carlos\\ Calle Tulip\'an s/n\\ M\'ostoles, Spain\\ ana.gomez.perez@urjc.es}
\and
\IEEEauthorblockN{Domingo G\'omez-P\'erez}
\IEEEauthorblockA{Universidad de Cantabria\\ Avenida de los Castros s/n\\ Santander, Spain\\ domingo.gomez@unican.es}
}

\maketitle

\begin{abstract}
The key agreements that protect a Wi-Fi connection rely on Diffie--Hellman, which a sufficiently capable quantum computer would break; an adversary can therefore record an association today and decrypt its traffic years later. Data encryption itself is far less affected, so key establishment is the part of IEEE 802.11 that has to be migrated first. This work targets the two protocols that perform that agreement inside the frames used to set up the link: Opportunistic Wireless Encryption (OWE), which encrypts otherwise open networks, and Fast Initial Link Setup (FILS), designed for fast connection in dense and mobile deployments. We propose replacing their Diffie--Hellman exchange with a hybrid one that runs the classical elliptic-curve agreement alongside ML-KEM, the key-encapsulation mechanism standardized by NIST (FIPS 203), and derives the master key from both secrets, so the connection stays protected as long as either component remains unbroken. 
The exchange reuses the frames each protocol already sends and adds no extra round trip; an access point can advertise the hybrid mode alongside classical OWE so that both coexist during the transition, and a station configured to require it refuses to associate without a completed post-quantum exchange; binding the advertised capabilities to the derived key, which would make tampering detectable, is specified as a design extension and is not part of the prototype. Rather than estimating costs on paper, we validate the design with a working implementation on the reference open-source Wi-Fi stack (hostapd and wpa\_supplicant): both endpoints derive the same key in every measured connection, with the two smaller parameter sets each enlarged association frame still fits in a single frame even under a conservative 1500-byte fragmentation threshold, and the added connection time and computation stay below half a millisecond. The measurements are obtained on OWE; the design carries over to FILS, whose frame budget is left for specific measurement. Of the three ML-KEM parameter sets, the intermediate one offers the best balance between security level and size.

\end{abstract}
\vspace{6pt}

\section{Introduction}
The security of modern wireless networks rests on public-key cryptographic primitives whose strength is based on the hardness of the factoring and discrete-logarithm problems. Shor's algorithm, run on a cryptographically relevant quantum computer, solves both problems efficiently~\cite{shor1997}, while Grover's algorithm halves the effective strength of symmetric and hash primitives~\cite{grover1996}. The direct consequence for IEEE 802.11 is that AES-based data encryption remains essentially secure when the key size is raised to 256 bits, but the key-establishment and authentication mechanisms become exposed.

This asymmetry is the crux of the problem. An adversary with access to a quantum computer does not need to break traffic encryption directly: it suffices to compromise the key agreement to derive the session key. Moreover, the \emph{harvest-now-decrypt-later} strategy allows capturing association traffic today and decrypting it in the future, which brings the urgency of migration forward to before real quantum hardware becomes  available~\cite{tan2023}. Mosca's theorem formalizes this urgency: if the time during which the data must remain secret ($X$) plus the time needed to deploy a solution ($Y$) exceeds the time until a relevant quantum computer appears ($Z$), then it is already too late to start~\cite{mosca2018}.

Within IEEE 802.11, a quantum adversary compromises every key agreement and signature in use: SAE, OWE, FILS, AP PeerKey, PASN, and the EAP methods, EAP-TLS among them~\cite{harkins2024}. This work addresses two of them: Opportunistic Wireless Encryption (OWE) and Fast Initial Link Setup (FILS). Both can establish an ephemeral Diffie--Hellman secret inside the management frames that set up the link---OWE in the (re)association frames and FILS in the authentication frames---so the agreement can be replaced by a key-encapsulation mechanism (KEM) operating on the same frames and without altering the number of exchanges. They also matter operationally: FILS targets the dense, mobile, and IoT deployments least tolerant of additional bytes per association, and OWE is what gives open networks the confidentiality they otherwise lack~\cite{rfc8110}. SAE and EAP-TLS fall outside this scope for opposite reasons: SAE is a password-authenticated key exchange (PAKE), for which no KEM is a drop-in substitute and whose post-quantum counterpart remains an open design problem; EAP-TLS, in turn, inherits the hybrid handshake specified by the IETF in RFC~9954~\cite{stebila2024}.

Within FILS, the authentication variant determines whether the threat applies at all: Shared Key authentication derives its keys from a symmetric EAP re-authentication context, beyond the reach of Shor's algorithm, unless its perfect-forward-secrecy option is enabled, whereas Public Key authentication always performs the exchange~\cite{fils2016}. The migration proposed here applies to the variants that perform it, whose differing frame budgets are examined in Section~\ref{sec:fils}.

The contributions of this paper are as follows:
\begin{enumerate}
\item A migration of the OWE and FILS key agreements to a hybrid ECDH\,$+$\,ML-KEM-768 scheme, with the message flow adapted to the asymmetric semantics of a KEM and placed in the frames each protocol already exchanges: the association frames in OWE and the authentication frames in FILS.

\item A crypto-agile negotiation design that lets an access point offer the classical and the hybrid agreement simultaneously during the transition. The prototype signals the hybrid capability with an experimental element and implements a local policy, on the AP and on the station, that refuses a classical fallback; binding the advertised suites to the derived key, so that tampering is detected once the pairwise master key (PMK) is established, is specified but not implemented.

\item A reference implementation on \textit{hostapd} and \textit{wpa\_supplicant} using ML-KEM (FIPS 203) and ECDH as provided by a production cryptographic library, which verifies that both endpoints derive an identical PMK and measures the cost the migration introduces: frame sizes with MAC headers, fragmentation behaviour, connection setup latency, and the computational cost of the primitives.

\item A security argument for the key-derivation combiner, together with an empirical verification that the implementation meets the operational preconditions on which forward secrecy depends. The measurements are obtained on OWE; the cryptographic and computational core carries over to FILS, whose frame budget is examined in Section~\ref{sec:fils}.
\end{enumerate}

The remainder of the paper is organized as follows. Section~\ref{sec:sota} reviews the state of the art, covering the quantum threat model, the standardization of post-quantum algorithms, and the prior proposals for migrating wireless networks, against which the contribution of this work is delimited. Section~\ref{sec:fund} provides the background needed to follow the proposal: where OWE and FILS place their key agreement within the 802.11 exchange, and how the asymmetric semantics of a KEM differ from the contributory semantics of Diffie--Hellman, a difference that determines which endpoint plays which role in the migrated flow. Section~\ref{sec:prop} describes the proposal itself---the hybrid message flow, the combiner that derives the PMK from both secrets, and the crypto-agile negotiation---and Section~\ref{sec:eval} validates it experimentally on \textit{hostapd} and \textit{wpa\_supplicant}, reporting frame sizes, fragmentation, latency, and computational cost, and discussing what these results imply for FILS and what they leave open. Section~\ref{sec:concl} concludes and outlines future work.

\section{State of the Art}\label{sec:sota}

Tan \emph{et al.}~\cite{tan2023} develop a post-quantum adversarial model from the user's perspective, identifying adversarial personas and target prioritization. Their analysis establishes that adversaries will attack asymmetric-key applications first, the most exposed to Shor's algorithm, and recommends early mitigations, among them perfect forward secrecy and symmetric keys of at least 256 bits. The primitives needed to deploy those mitigations are now standardized: NIST finalized FIPS 203 in 2024~\cite{fips203}, specifying ML-KEM, a lattice-based key-encapsulation mechanism whose security rests on the hardness of the Module Learning with Errors problem. Compared with the post-quantum signature standards released alongside it, ML-KEM offers keys and ciphertexts of moderate size, which makes it the natural candidate for key establishment in media constrained by frame size.

Harkins~\cite{harkins2024} sets out the problem for 802.11 and proposes replacing Diffie--Hellman with ML-KEM in FILS and in OWE, noting that SAE requires a post-quantum PAKE and that the 4-way handshake should be reinforced with SHA-3. This line of work has since matured into a formal task group, P802.11bt, with approved PAR and CSD and a draft under ballot~\cite{tgbt}.\footnote{At the time of writing, draft D1.0 has undergone its first Working Group letter ballot (LB297, closed September 2026).} Relevantly here, the PAR scope covers AKM suites for post-quantum cryptography, key-establishment and digital-signature algorithms that use it, a post-quantum password-authenticated key exchange, and modifications to the key handshake protocols; replacing the Diffie--Hellman agreement of OWE and FILS falls under the key-establishment item, which places this proposal within the scope of the amendment under development. Keh and Chen~\cite{keh2026} take a different route: instead of modifying 802.11, they add an AES encryption layer on top of WPA whose key is established via ML-KEM. It is a pragmatic solution for
industrial environments and legacy devices, but it acts as a compensating control and leaves the underlying protocol unchanged. NIST's guidance on crypto agility~\cite{nistcswp39} provides strategies and practices for replacing algorithms without disrupting operations, although without Wi-Fi specificity.

What is missing from this body of work is a protocol-level design for the OWE and FILS key agreement over hybrid ML-KEM that has been built and measured rather than only specified. Harkins~\cite{harkins2024}, the closest prior work, proposes the substitution conceptually but provides no implementation, no precise placement within each protocol's
state machine, and no measurements. This paper does not contribute a new cryptographic construction---the building blocks, hybrid ECDH\,$+$\,ML-KEM, a KDF-based combiner, and encapsulation in existing frames, deliberately rely on known and analyzed constructions~\cite{bindel2019,stebila2024}---but an engineering and empirical characterization: to the best of our knowledge, the first public implementation of hybrid PQ-OWE on \textit{hostapd} and
\textit{wpa\_supplicant} with real ML-KEM; the placement of the agreement in OWE's association frames rather than its authentication frames, with a role assignment that preserves the preconditions for forward secrecy; and the measurement of the resulting feasibility envelope---frame sizes with MAC headers, fragmentation threshold, connection setup latency, and computational cost---that analytical treatments leave open.

\section{Background}\label{sec:fund}
\subsection{Key establishment in OWE and FILS}\label{sec:keyest}

OWE~\cite{rfc8110} and FILS~\cite{fils2016} differ in where the agreement is carried. OWE uses Open System authentication and performs the Diffie--Hellman agreement in the \emph{(Re)Association} frames: the station (STA) sends its public value in the DH Parameter element of the \emph{Association Request}, and the access point (AP) answers in the \emph{Association Response}. FILS instead carries the ephemeral agreement in the two Authentication frames that precede association, \emph{Auth seq~1} (STA$\to$AP) and \emph{Auth seq~2} (AP$\to$STA), and folds key confirmation into the (Re)Association exchange. In both cases the resulting secret feeds the derivation of the pairwise master key (PMK), from which the pairwise transient key (PTK) is obtained---through the 4-way handshake in OWE, and within the association exchange itself in FILS.

\subsection{KEM versus non-interactive key exchange}
The structural difference between a KEM and Diffie--Hellman conditions the design. A non-interactive key exchange (NIKE), of which Diffie--Hellman is the canonical instance, is a pair of algorithms $(\mathsf{KeyGen},\mathsf{SharedKey})$: each party runs $(sk_i,pk_i)\leftarrow\mathsf{KeyGen}()$, publishes $pk_i$, and computes $z=\mathsf{SharedKey}(sk_i,pk_j)$, with correctness requiring $\mathsf{SharedKey}(sk_A,pk_B)=\mathsf{SharedKey}(sk_B,pk_A)$. The exchange is contributory and its messages are independent, so either party may send first. A KEM is a triple $(\mathsf{KeyGen},\mathsf{Encaps},\mathsf{Decaps})$: the initiator runs $(ek,dk)\leftarrow\mathsf{KeyGen}()$ and sends the encapsulation key $ek$; the responder runs $(ct,z)\leftarrow\mathsf{Encaps}(ek)$ and returns the ciphertext $ct$; the initiator recovers $z=\mathsf{Decaps}(dk,ct)$ with the decapsulation key $dk$. The second message depends on the first, which fixes the roles of the two endpoints. Throughout the paper, $z_{cl}$ denotes the classical (ECDH) shared secret and $z_{pq}$ the post-quantum (ML-KEM) one.

This semantics fits a two-frame exchange without additional round trips. When the ML-KEM key pair is generated ephemerally per association (by the STA) and the decapsulation key is securely destroyed after use, the scheme offers forward-secrecy properties comparable to those of an ephemeral Diffie--Hellman; this property does not derive from the algorithm itself, but from the ephemeral management of the keys. Formally, a key-establishment protocol provides forward secrecy if an adversary that records a session and later obtains all long-term secrets of both endpoints, as well as any state they retain after the session has ended, still cannot distinguish that session's key from random. The property is therefore relative to an adversary model and cannot be proven by inspecting code; what an implementation can verify are its operational preconditions (Section~\ref{sec:secmodel}).
Table~\ref{tab:mlkem} summarizes the ML-KEM parameter sizes.

\begin{table}[!tb]
\caption{ML-KEM parameter sizes (bytes)}
\label{tab:mlkem}
\centering
\begin{tabular}{lrrrr}
\toprule
Variant & Enc. key & Dec. key & Cipher. & Secret \\
\midrule
ML-KEM-512  & 800  & 1632 & 768  & 32 \\
ML-KEM-768  & 1184 & 2400 & 1088 & 32 \\
ML-KEM-1024 & 1568 & 3168 & 1568 & 32 \\
\bottomrule
\end{tabular}
\end{table}

\section{Migration Proposal}\label{sec:prop}
\subsection{Hybrid message flow}
We propose replacing the Diffie--Hellman agreement of OWE with a hybrid exchange that combines ECDH over OWE's default P-256 group (the prototype supports groups 19--21) and ML-KEM-768. Following the KEM semantics and aiming to preserve per-association forward secrecy, \emph{the STA is the one that generates the ephemeral ML-KEM key pair}. In the OWE case: in the \emph{Association Request} frame the STA sends its public ECDH value and its ML-KEM encapsulation key; in the \emph{Association Response} the AP replies with its public ECDH value and with the ML-KEM ciphertext resulting from encapsulating against the received key. In FILS, the same exchange is located in the two authentication frames. Each endpoint thus obtains two shared secrets, $z_{cl}$ (ECDH) and $z_{pq}$ (ML-KEM).

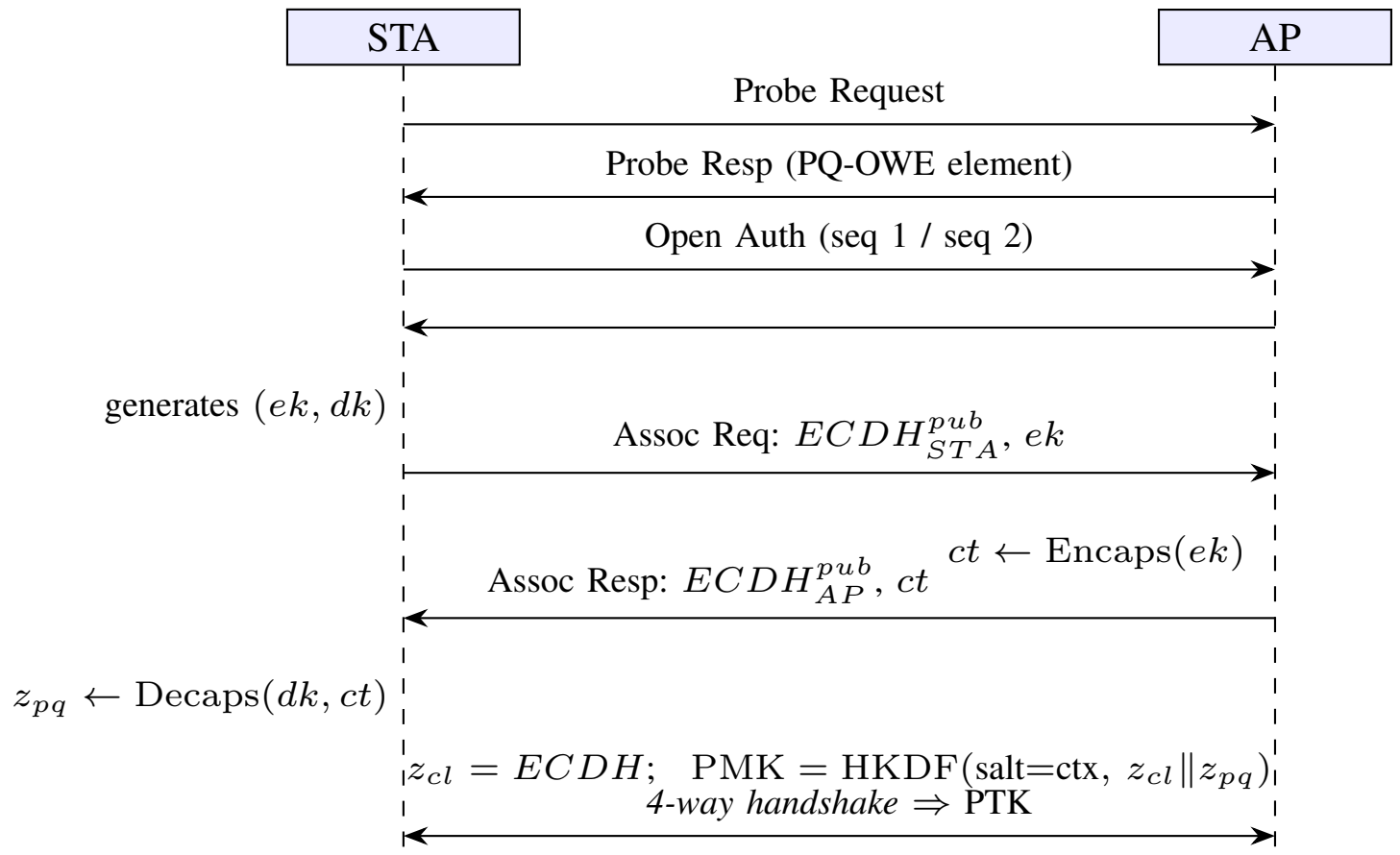
\begin{figure}[!tb]
\centering
\footnotesize
\begin{tikzpicture}[>=Stealth,every node/.style={font=\footnotesize}]
  \def\xs{0}    \def\xa{6.0}   \def\ytop{0}   \def\ybot{-5.6}
  \node[draw,fill=blue!8,minimum width=1.6cm] (sta) at (\xs,\ytop) {STA};
  \node[draw,fill=blue!8,minimum width=1.6cm] (ap)  at (\xa,\ytop) {AP};
  \draw[dashed] (sta) -- (\xs,\ybot);
  \draw[dashed] (ap)  -- (\xa,\ybot);
  \draw[->] (\xs,-0.6) -- node[above,midway]{\scriptsize Probe Request} (\xa,-0.6);
  \draw[<-] (\xs,-1.1) -- node[above,midway]{\scriptsize Probe Resp (PQ-OWE element)} (\xa,-1.1);
  \draw[->] (\xs,-1.6) -- node[above,midway]{\scriptsize Open Auth (seq 1 / seq 2)} (\xa,-1.6);
  \draw[<-] (\xs,-2.0) -- (\xa,-2.0);
  \node[left=0pt,align=left] at (\xs,-2.55) {\scriptsize generates $(ek,dk)$};
  \draw[->] (\xs,-3.0) -- node[above,midway]{\scriptsize Assoc Req: $ECDH^{pub}_{STA}$, $ek$} (\xa,-3.0);
  \node[anchor=east] at (5.9,-3.55) {\scriptsize $ct \leftarrow \mathrm{Encaps}(ek)$};
  \draw[<-] (\xs,-4.0) -- node[above,pos=0.35]{\scriptsize Assoc Resp: $ECDH^{pub}_{AP}$, $ct$} (\xa,-4.0);
  \node[left=0pt,align=left] at (\xs,-4.55) {\scriptsize $z_{pq}\leftarrow \mathrm{Decaps}(dk,ct)$};
  \node[align=center] at (3,-5.05) {\scriptsize $z_{cl}=ECDH$; \; $\mathrm{PMK}=\mathrm{HKDF}(\text{salt}{=}\text{ctx},\,z_{cl}\|z_{pq})$};
  \draw[<->] (\xs,-5.5) -- node[above,midway]{\scriptsize \emph{4-way handshake} $\Rightarrow$ PTK} (\xa,-5.5);
\end{tikzpicture}
\caption{Message flow of hybrid PQ-OWE. The STA generates the ephemeral ML-KEM key pair and sends $ek$ in the \emph{Association Request}; the AP encapsulates and returns $ct$ in the \emph{Association Response}. No round trips are added over classical OWE. In FILS the same exchange is located in the authentication frames.}
\label{fig:proto}
\end{figure}

Fig.~\ref{fig:proto} summarizes the flow for OWE. This split---the STA's encapsulation key in the first frame, the AP's ciphertext in the second---allows resolving the agreement in a single request/response round without pre-association key material and without long-lived key state at the AP.

\subsection{KDF-based hybrid combiner}
Both secrets are combined via the HKDF construction of RFC 8110, extended with the post-quantum component. The classical and post-quantum secrets form the Input Keying Material (IKM), while the public session context acts as the salt of the extraction:
\begin{equation}
\label{eq:hkdf}
\begin{aligned}
\mathrm{prk} &= \text{HKDF-Extract}\big(\text{salt}=\text{ctx},\; \text{IKM}=z_{cl} \,\|\, z_{pq}\big),\\
\mathrm{PMK} &= \text{HKDF-Expand}\big(\mathrm{prk},\; \text{``OWE Key Generation''}\big),
\end{aligned}
\end{equation}
where $\text{ctx}=C \,\|\, A \,\|\, \text{group} \,\|\, \text{ct}$ binds the derivation to the session context (STA and AP public ECDH keys, ECDH group identifier, and ML-KEM ciphertext); $ct$ binds the PMK to the specific ML-KEM exchange. In the prototype the AKM and the RSNE are not part of $\text{ctx}$. Placing the secrets in the IKM and the public context in the salt is deliberate: it keeps the combiner's min-entropy argument clean, since the min-entropy of the IKM conditioned on the adversary's view is lower-bounded by that of the surviving component, and the non-secret salt only improves the extraction.

\emph{Security model and goal.} We consider a probabilistic polynomial-time (PPT) adversary with classical capability, active network access, and future access to a quantum computer able to run Shor's algorithm, and hence to solve the discrete logarithm problem in the elliptic-curve and finite-field groups used by OWE. The goal of the combiner is PMK indistinguishability: the PMK is indistinguishable from random as long as at least one of the two concatenated input secrets ($z_{cl}$ or $z_{pq}$) remains indistinguishable from random to the adversary.

The construction of Eq.~\eqref{eq:hkdf} ---concatenation of secrets followed by HKDF extraction--- is an instance of the hybrid-KEM combiner whose security preservation under the dual-PRF assumption of HMAC was proven by Bindel \emph{et al.}~\cite{bindel2019} and adopted in the hybrid design of TLS 1.3~\cite{stebila2024}. Both secrets have a fixed length for a given suite (the ECDH secret is left-padded to the field size, as in RFC~8110, and $z_{pq}$ is always 32 bytes), which is the condition under which that design specifies the concatenation.

\subsection{Crypto-agile negotiation and downgrade mitigation}\label{sec:negot}
In the proposed design, the post-quantum capability is advertised via new Authentication and Key Management (AKM) suite selectors in the Robust Security Network Element (RSNE) and, in FILS, new authentication algorithm numbers. In the prototype it is signalled instead by an experimental Extension element (Extension ID 200, one octet identifying the ML-KEM variant) carried only in the Probe Response, not in the Beacon, and the AKM remains OWE.

A transitioning AP serves classical and hybrid stations simultaneously: a station that understands the hybrid capability uses the ML-KEM variant advertised by the AP, while a legacy station ignores it and runs classical OWE. The main risk is a downgrade attack. An active adversary that rewrites the Beacon and Probe Response frames strips the hybrid capability (the AKM selector in the proposed design, the experimental element in the prototype); the STA then runs classical OWE with the genuine AP, and the adversary only needs to record the exchange and decrypt it once a quantum computer is available. This is cheaper than a full man-in-the-middle, which must relay the whole session in real time and against which OWE offers no protection by design, since it lacks identity authentication.

We propose three countermeasures as part of the design; the first two are not implemented in the prototype, where, with the default configuration, this downgrade goes undetected. First, the AP's complete advertised RSNE (all offered AKMs, not only the selected one) is added to $\text{ctx}$, so that the PMK depends on what the AP actually offered; the presence of $ct$ in $\text{ctx}$ already binds the PMK to that specific ML-KEM exchange (\emph{transcript binding}). Second, as in WPA2/WPA3, the AP repeats its RSNE in message~3 of the 4-way handshake, protected by the MIC under a key derived from the PMK; the STA compares it with the RSNE received before association and aborts on mismatch. An adversary that only rewrote the beacons does not know the PMK and cannot forge that MIC, so the downgrade is detected. This check is effective only once the hybrid capability is encoded in the RSNE, as proposed: in the prototype the RSNE is identical in both modes. Third, once a STA has associated with an SSID in hybrid mode, it can record that capability and refuse later classical associations to the same network, analogous to the WPA3 \emph{Transition Disable} indication. The prototype implements a configured, rather than learned, form of this third mechanism: a policy flag (\texttt{owe\_pqc\_required}) available on both endpoints. On the AP it rejects Association Requests without a valid encapsulation key, at the price of excluding classical stations; on the station it suppresses the Association Request when the AP advertises no ML-KEM variant and rejects an Association Response without a valid ciphertext. On the testbed, a station with the policy enabled refused 6/6 associations with a classical-only AP and completed 6/6 with a hybrid one. The policy is local and is not bound to the PMK. The scope of this protection must be stated precisely: the first two mechanisms detect tampering \emph{after} PMK establishment, and the third relies on trust on first use or, in its configured form, on the station knowing in advance that the network is hybrid-capable; otherwise none of them closes the downgrade \emph{before} the first association, since beacons and probes are unprotected at that point.

Because the prototype carries the capability only in Probe Responses, the same fallback can also occur without an adversary. A station that learns the BSS from Beacons alone runs classical OWE. On the testbed this happened in \nfallback{} of the \nfallbackTot{} connection attempts made against a hybrid-capable AP: the AP did send the Probe Response, but 35\,ms after the request (at most 18\,ms in the other 32 connections of that run), which is consistent with the station having already left the channel. Advertising the capability in Beacons as well, or enabling the station policy, removes this benign fallback.

\section{Experimental Validation and Discussion}\label{sec:eval}
Rather than estimating sizes and costs analytically, the figures in this section come from a reference implementation of the proposed exchange on \textit{hostapd}~2.10 (AP) and \textit{wpa\_supplicant} (STA), with  ML-KEM (FIPS 203) and ECDH provided by OpenSSL\footnote{Native ML-KEM implementation of the OpenSSL default provider, available since version 3.5 (\texttt{EVP\_PKEY-ML-KEM}, \url{https://docs.openssl.org/3.5/man7/EVP_PKEY-ML-KEM/}); no external provider such as liboqs is used.} and an HKDF combiner per Eq.~\eqref{eq:hkdf} that, as in RFC~8110, uses the hash of the ECDH group (SHA-256 for P-256, the group used in all measurements).

The implementation exercises the hybrid flow for OWE and verifies, on each connection, that the STA and AP derive an identical PMK (30/30 connections per mode, $N=30$ after 3 warmup) and that the subsequent data transfer completes (30/30); Fig.~\ref{fig:evidence} summarizes that output. It also measures an in-process cryptographic microbenchmark (\texttt{CLOCK\_MONOTONIC}) for the three ML-KEM variants. The functional validation is performed on the user-space testbed; physical-layer airtime and throughput figures are not reported as RF measurements (see Section~\ref{sec:concl}).

\subsection{Experimental environment}
For reproducibility, measurements were obtained on: \envCPU, \envRAM, \envOS, with \envOpenSSL{} (native ML-KEM provider), \envCC{}, and \envHostapd{} on the \envDriver{} driver. Throughout this section, OWE denotes classical OWE (ECDH only) and PQ-OWE the hybrid ECDH\,$+$\,ML-KEM-768 scheme. Latency (Phase~A) uses $N=30$ connections per mode after 3 warmup. The two modes were measured in consecutive blocks of 33 connections each (classical OWE first), so a drift of the host over time cannot be excluded as a confounder. $\Delta t_{\text{assoc}}$ is the interval between the Association Request and the Association Response as seen by a monitor interface; it covers the AP's ECDH, encapsulation and frame construction, but not the STA's key generation and decapsulation, which fall outside it. The exchange time is the sum of three such intervals (authentication, association and 4-way handshake).

The ML-KEM-512 and ML-KEM-1024 frame measurements, the interleaved latency replication, the validation of the station policy and the microbenchmark of Table~\ref{tab:micro} were obtained later on a second host: Intel Core Ultra~7 258V, 2 vCPUs in a VirtualBox VM, 3.8\,GiB, kernel 6.18.12, GCC 15.2.0 and the same OpenSSL version. Frame sizes and fragmentation do not depend on the host; absolute timings do, and are not compared across hosts.

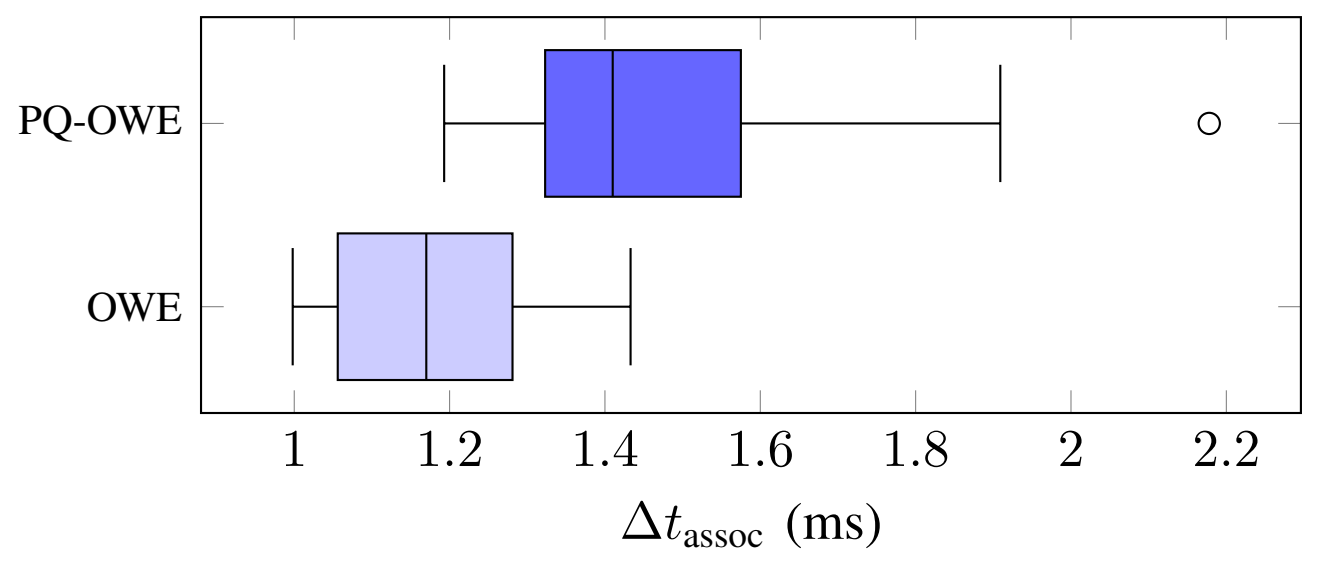
\begin{figure}[!tb]
\centering
\begin{tikzpicture}
\begin{axis}[
  width=\columnwidth, height=4.2cm,
  boxplot/draw direction=x,
  xlabel={$\Delta t_{\text{assoc}}$ (ms)},
  ytick={1,2}, yticklabels={OWE, PQ-OWE},
  yticklabel style={font=\footnotesize},
]
\addplot+[boxplot prepared={lower whisker=\bpOWElw, lower quartile=\bpOWEq, median=\dtassocOWE, upper quartile=\bpOWEQ, upper whisker=\bpOWEuw}, fill=blue!20, draw=black] coordinates {};
\addplot+[boxplot prepared={lower whisker=\bpPQlw, lower quartile=\bpPQq, median=\dtassocPQC, upper quartile=\bpPQQ, upper whisker=\bpPQuw}, fill=blue!60, draw=black, mark=o, mark options={draw=black}] coordinates {(0,\bpPQout)};
\end{axis}
\end{tikzpicture}
\caption{Distribution of the association latency $\Delta t_{\text{assoc}}$ per mode ($N=\nlat$ connections each, block-ordered run). Boxes show the IQR, the line the median, whiskers the most extreme value within 1.5$\times$IQR, and the circle an outlier.}
\label{fig:latdist}
\end{figure}

The microbenchmark (Phase~B) uses $N=\nmicro$ iterations per operation, discarding the first five (cold cache). The fragmentation sweep (Phase~C) uses 5 repetitions per threshold. Medians and IQR are reported. A non-parametric test was chosen a priori: association latencies on a shared host are bounded below by the protocol cost and right-skewed by scheduler jitter, so normality cannot be assumed. Significance is therefore assessed with the Mann--Whitney $U$ test; Fig.~\ref{fig:latdist} shows both distributions. The patch code and measurement scripts will be released for verification.

\begin{figure}[!tb]
\centering
\footnotesize
\begin{verbatim}
OWE vs OWE+ML-KEM-768  (N=30, failures 0/0)
                      OWE      PQ-OWE
AssocReq  (B)         223      1419
AssocResp (B)         206      1306
Assoc dt  (ms, med)   1.17     1.41   p=1.8e-7
pmk_match             30/30    30/30
transfer_ok           30/30    30/30
fragm@1500 (PQ-OWE)  AssocReq:no  AssocResp:no
\end{verbatim}
\caption{Summary of the testbed results on \textit{hostapd}/\textit{wpa\_supplicant}, condensed from the benchmark output: identical PMK derivation (\texttt{pmk\_match} 30/30) and correct transfer across the 30 connections per mode, frame sizes, and absence of fragmentation at the 1500 threshold.}
\label{fig:evidence}
\end{figure}

\subsection{Frame-size budget and fragmentation}
The main feasibility challenge is size. Table~\ref{tab:overhead} reports, for classical OWE and for each ML-KEM variant of the hybrid scheme, the association frames measured on \textit{hostapd} with MAC headers and Information Elements included. Sizes are given as captured, i.e., including the radiotap header that the monitor interface prepends (26 and 13\,B, respectively), so the 802.11 frames themselves are that much smaller; the overhead is unaffected. The table also reports the number of fragments obtained when sweeping the 802.11 fragmentation threshold. With ML-KEM-768, the STA transmits the encapsulation key (1184\,B) in the \emph{Association Request} and the AP the ciphertext (1088\,B) in the \emph{Association Response}, each carried in an Extension element with element fragmentation, which adds 12\,B (4\,B plus 2\,B per fragment); the measured overhead is 2296\,B per handshake ($ek + ct + 24$\,B of element headers). The decapsulation key (2400\,B in ML-KEM-768) is not transmitted: it remains at the STA. The resulting frames are \areqPQC\,B (Association Request) and \arespPQC\,B (Association Response), both below 1500 bytes. IEEE~802.11 does not fragment management frames by default (the fragmentation threshold is disabled in \textit{hostapd} unless configured, and a management frame body may reach 2304 bytes), so the sweep takes 1500 bytes, the Ethernet MTU, as a conservative reference. At that threshold neither frame is fragmented; fragmentation appears only at lower thresholds (2 fragments at 1024\,B, 3 at 512\,B), while classical OWE never fragments. This empirically validates the central feasibility claim: with ML-KEM-768 the migration fits in a single association frame in realistic configurations.

\begin{table}[!tb]
\caption{Measured association frame sizes (bytes, as captured: MAC header, Information Elements and radiotap header of 26/13\,B) and number of fragments of each frame (Request/Response) at three 802.11 fragmentation thresholds, for classical OWE and the hybrid scheme with each ML-KEM variant (\textit{hostapd} PoC). Overhead is the sum of both frames minus classical OWE.}
\label{tab:overhead}
\centering
\footnotesize
\setlength{\tabcolsep}{3pt}
\begin{tabular}{lcrrrccc}
\toprule
 & NIST & Assoc & Assoc & Overh. & \multicolumn{3}{c}{Fragments (Req/Resp)} \\
\cmidrule(lr){6-8}
Scheme & level & Req. & Resp. & (B) & 1500 & 1024 & 512 \\
\midrule
Classical OWE   & --- & \areqOWE & \arespOWE & ---    & 1/1 & 1/1 & 1/1 \\
+ ML-KEM-512    & 1   & \areqFiveOneTwo & \arespFiveOneTwo & \ovhFiveOneTwo & 1/1 & 1/1 & 3/2 \\
+ ML-KEM-768    & 3   & \areqPQC & \arespPQC & 2296   & 1/1 & 2/2 & 3/3 \\
+ ML-KEM-1024$^{*}$ & 5 & \areqTenTwoFour & \arespTenTwoFour & \ovhTenTwoFour & 2/2 & 2/2 & 4/4 \\
\bottomrule
\end{tabular}
\\[2pt]
\raggedright {\scriptsize $^{*}$With the supplicant's association-request buffer enlarged from 1500 to 2304\,B (Section~\ref{sec:variants}).}
\end{table}

\subsection{Comparison of ML-KEM variants}\label{sec:variants}
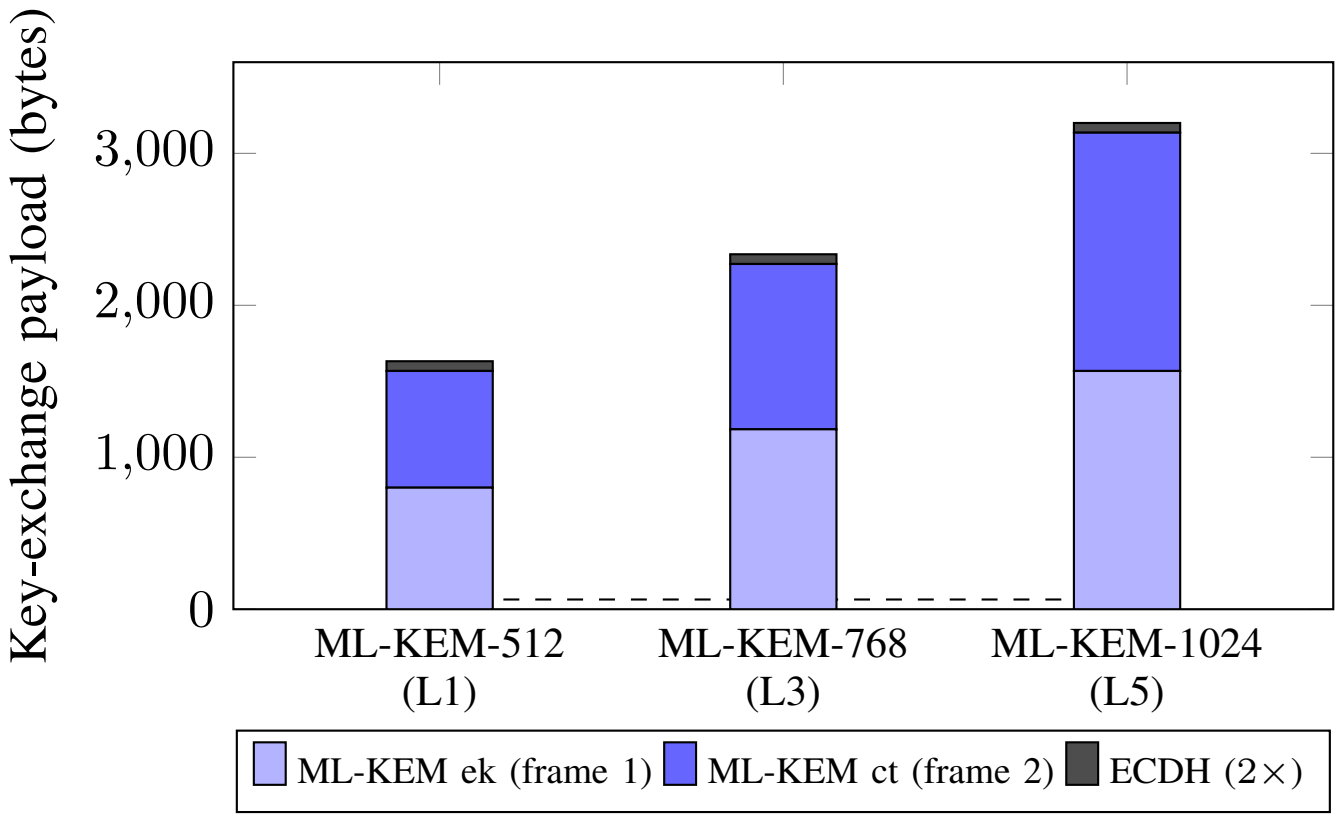
\begin{figure}[!tb]
\centering
\begin{tikzpicture}
\begin{axis}[
  width=\columnwidth, height=5.2cm,
  ybar stacked, bar width=20pt,
  ymin=0, ymax=3600,
  ylabel={Key-exchange payload (bytes)},
  symbolic x coords={512,768,1024},
  xtick=data,
  xticklabels={ML-KEM-512\\(L1),ML-KEM-768\\(L3),ML-KEM-1024\\(L5)},
  xticklabel style={align=center,font=\footnotesize},
  legend style={at={(0.5,-0.22)},anchor=north,legend columns=3,font=\scriptsize},
  enlarge x limits=0.3,
  nodes near coords align={vertical},
]
\addplot[fill=blue!30, draw=black] coordinates {(512,800) (768,1184) (1024,1568)};
\addplot[fill=blue!60, draw=black] coordinates {(512,768) (768,1088) (1024,1568)};
\addplot[fill=black!70, draw=black] coordinates {(512,64) (768,64) (1024,64)};
\legend{ML-KEM ek (frame 1), ML-KEM ct (frame 2), ECDH ($2\times$)}
\draw[dashed] (axis cs:512,64) -- (axis cs:1024,64);
\end{axis}
\end{tikzpicture}
\caption{Key-exchange payload per handshake of the three ML-KEM variants in the hybrid scheme, broken down into encapsulation key (frame 1), ciphertext (frame 2), and the two ECDH public values. The dashed line marks classical OWE (64\,B). Element headers are not included; the measured frame overhead is given in Table~\ref{tab:overhead}.}
\label{fig:overhead}
\end{figure}

The choice of variant involves a trade-off between security level and size. Table~\ref{tab:overhead} and Fig.~\ref{fig:overhead} summarize the comparison. ML-KEM-512 (NIST level 1) minimizes the public data exchanged (1632\,B) and fits in a single frame down to a 1024-byte fragmentation threshold. ML-KEM-1024 (level 5) raises the public data to 3200\,B. Its 1584-byte key element exceeds the 1500-byte buffer that \textit{wpa\_supplicant}~2.10 reserves for its own association-request elements, so an unmodified supplicant cannot associate with this variant; this is itself a deployment constraint. With the buffer enlarged to 2304\,B, both frames are already fragmented at the 1500-byte reference threshold, although they are sent unfragmented when no threshold is configured (1781 and 1777\,B without the radiotap header, below the 2304-byte limit). ML-KEM-768 (level 3) offers the recommended balance: 2336\,B of public data and a single frame down to a 1500-byte threshold.

\subsection{Comparison with alternatives}
Table~\ref{tab:compare} positions the proposal against the relevant alternatives. All KEM-based variants resolve the agreement without additional round trips; the differences lie in the payload, the hybrid security net, and the degree of validation. Pure ML-KEM (without a classical component) saves the 64\,B of ECDH but loses the guarantee of not degrading classical security should ML-KEM be broken, so it is discouraged during transition. Harkins's proposal~\cite{harkins2024} is conceptually close but is formulated as a tech report with no implementation or measurements, and without a precise placement of the exchange within each protocol.

\begin{table}[!tb]
\caption{Comparison with alternatives (post-quantum OWE/FILS). Payload is the key-exchange material carried by the two frames (public values, encapsulation key and ciphertext), without element headers. All schemes resolve the agreement in two frames (0 extra RTT).}
\label{tab:compare}
\centering
\footnotesize
\begin{tabular}{lccc}
\toprule
Scheme & Payload & Hybrid & Impl.\\
       & (B) & sec. & $+$meas.\\
\midrule
Classical OWE (ECDH)         & 64   & ---  & standard \\
Pure ML-KEM (no classical)   & 2272 & no   & no \\
Harkins (FILS/OWE)~\cite{harkins2024} & $\sim$2300 & yes$^{*}$ & no \\
\textbf{PQ-OWE (this work)}  & \textbf{2336} & \textbf{yes} & \textbf{yes} \\
\bottomrule
\end{tabular}
\\[2pt]
\raggedright {\scriptsize $^{*}$Proposed conceptually; no published implementation or empirical characterization.}
\end{table}

\subsection{Transmission overhead and latency}
The additional cost of the proposal is transmission. Analytically, for ML-KEM-768 the $\sim$2336\,B of payload entail an airtime of about 3.1\,ms at 6\,Mbps, the lowest OFDM rate and one at which management frames are commonly sent, and about 18.7\,ms at the 1\,Mbps rate still used for them in the 2.4\,GHz band; at higher rates the figure shrinks proportionally ($\sim$346\,\textmu s at 54\,Mbps). Empirically, in a block-ordered run on the testbed the hybrid mode added a median \dtassocDelta\,ms to the association exchange: the median of $\Delta t_{\text{assoc}}$ goes from \dtassocOWE\,ms to \dtassocPQC\,ms ($+$\dtassocPct\,\%; Mann--Whitney $U$, $p=\mwupval$, $N=\nlat$), and the summed exchange time from $\sim$\wcOWE\,ms to $\sim$\wcPQC\,ms. An interleaved replication on a different, noisier host found no statistically significant difference (medians 1.93 and 2.06\,ms, $p=0.28$, $N=\nlat$): there the effect is smaller than the run-to-run spread, whose interquartile range exceeds 1\,ms. Since \envDriver{} delivers frames in memory without real airtime, this difference reflects host-side processing (cryptographic computation and the construction and copying of the larger frames) rather than radio transmission; on real hardware, the airtime estimated above must be added to it. The extra cost is marginal on conventional clients and APs, but should not be assumed negligible in all scenarios: on IoT nodes with constrained radios and CPUs, on battery-powered sensors, or in massive-roaming scenarios where association is repeated with high frequency, both the transmission energy of the $\sim$2.3\,kB and its impact at low PHY rates require a specific evaluation ---especially sensitive in FILS, whose purpose is to minimize association latency.

\subsection{Computational cost}
ML-KEM operations are arithmetically light, being based on lattice operations. The in-process microbenchmark quantifies this point on the code paths used by the implementation (Table~\ref{tab:micro}). A complete ML-KEM-768 operation (keygen $+$ encaps $+$ decaps $+$ combiner) sums to \ttotSevenSix\,\textmu s, i.e., $\sim$0.19\,ms, about 5\% of the summed exchange time of about 4\,ms measured on the same host (of these operations, only the AP's encapsulation and combiner fall inside the measured intervals). The cost scales monotonically with the variant ($512 < 768 < 1024$), and the combiner itself takes about 4\,\textmu s. For reference, one P-256 ECDH operation (key generation and derivation) costs about 100\,\textmu s per endpoint, so the classical component is of the same order as the whole ML-KEM-768 exchange. The high percentiles (p99 of a few hundred \textmu s) correspond to operating-system scheduler jitter in the virtual machine, not to the primitive. On embedded microcontrollers (ARM Cortex-M4) the optimized implementations take about 0.4 to 1.1 million cycles per operation, depending on the variant, according to reference benchmarks~\cite{pqm4}, i.e., a few milliseconds per operation at a core clock of 168\,MHz; in the usual client and AP scenarios the limiting factor remains transmission, not computation.

\begin{table}[H]
\caption{In-process cryptographic microbenchmark (\textmu s, medians, OpenSSL, second host). Encaps includes the import of the raw encapsulation key, as done by the AP; the combiner is Eq.~\eqref{eq:hkdf}. Totals exclude ECDH (about 100\,\textmu s per endpoint for P-256).}
\label{tab:micro}
\centering
\begin{tabular}{lrrrrr}
\toprule
Variant & keygen & encaps & decaps & combiner & Total \\
\midrule
ML-KEM-512  & \tkgFiveOneTwo & \tenFiveOneTwo & \tdeFiveOneTwo & \thkFiveOneTwo & \textbf{\ttotFiveOneTwo} \\
ML-KEM-768  & \tkgSevenSix   & \tenSevenSix   & \tdeSevenSix   & \thkSevenSix   & \textbf{\ttotSevenSix} \\
ML-KEM-1024 & \tkgTenTwoFour & \tenTenTwoFour & \tdeTenTwoFour & \thkTenTwoFour & \textbf{\ttotTenTwoFour} \\
\bottomrule
\end{tabular}
\end{table}

\subsection{Security model and forward secrecy}\label{sec:secmodel}
Under the adversary model and the dual-PRF assumption of the combiner (Section~\ref{sec:prop}), the PMK remains indistinguishable from random as long as at least one of the two key-establishment mechanisms ---the classical component (ECDH) or the post-quantum one (ML-KEM)--- remains secure. The classical secret is always concatenated first and an ECDH failure aborts the association, so the PMK never depends on ML-KEM alone. Conversely, the prototype has no clean fallback: if the ML-KEM element is absent from the Probe Response or the Association Request, both ends silently derive a classical PMK (the downgrade of Section~\ref{sec:negot}); if the ciphertext is removed, malformed or altered (ML-KEM's implicit rejection), the two PMKs differ and the 4-way handshake fails. These fault cases follow from the code and were reproduced on a model of the implemented paths with fault injection; they were not exercised by tampering with frames on the testbed. With \texttt{owe\_pqc\_required} enabled, the silent cases become association failures.

What we verify in the implementation are the operational preconditions for forward secrecy (Section~\ref{sec:fund}): the ML-KEM key pair is generated ephemerally per association at the STA; the decapsulation key is generated inside the cryptographic library and held in an opaque key object (\texttt{EVP\_PKEY}) that is never exported to application memory; and it is zeroized by the library's key destructor as soon as the Association Response has been processed, on every exit path of that processing (success, failure, absence of the element, or PMKSA caching); a key pair whose response never arrives is discarded at the next association attempt or when the supplicant state is torn down. On both endpoints, $z_{pq}$ and the combined secret are cleared as soon as they have been used, the intermediate key with a zeroization the compiler cannot optimize away, and the PTK by the standard key handling on disconnection. The PMK is the exception: as in classical OWE, both endpoints keep it in the PMKSA cache (12 hours by default in this stack) so that a station can re-associate without a new exchange. Hence, once that cache entry has expired or been flushed, no state that would allow a recorded session to be decrypted survives---the same ephemeral-key model on which TLS~1.3 bases forward secrecy for its (EC)DHE and hybrid key exchanges~\cite{stebila2024}. Nothing more is claimed: like classical OWE, PQ-OWE lacks identity authentication and does not protect against an active man-in-the-middle; it restores confidentiality against a passive observer with quantum capability.

\subsection{Applicability to FILS}\label{sec:fils}
The experimental validation above was performed on OWE. It is worth specifying which results carry over to FILS and which require specific measurement, given the structural difference between the two: whereas OWE performs the agreement in the \emph{association frames} over Open System authentication, FILS places it in the \emph{authentication frames} (Auth seq~1 and seq~2, Section~\ref{sec:keyest}) and folds the key confirmation into the association itself, removing the separate 4-way handshake to minimize latency.

The KDF-based hybrid combiner and its security argument (Section~\ref{sec:prop}), the forward-secrecy preconditions (Section~\ref{sec:secmodel}), the computational cost of Table~\ref{tab:micro}, which measures the ML-KEM primitive rather than the protocol, and the post-quantum data overhead in bytes ($ek+ct$) are protocol-agnostic and carry over directly to FILS.

What does require specific measurement is the frame budget and its fragmentation. The result that the hybrid frame fits without fragmentation at 1500\,B was obtained on OWE's \emph{association frames}, whose baseline is small (\areqOWE/\arespOWE\,B). FILS authentication frames start from a larger, variant-dependent baseline: the \emph{Shared Key} variant carries the \emph{FILS Wrapped Data} (EAP-RP packet) and the \emph{Public Key} variant may carry a certificate chain of hundreds of bytes to several kB. Adding the 1184\,B of the encapsulation key on top of that baseline may cross the 1500\,B threshold ---especially in \emph{Public Key}, the most critical case---, so FILS fragmentation behavior cannot be inferred from OWE's and must be re-measured. Likewise, since FILS is a flow designed to minimize association latency, the same absolute transmission cost of the $\sim$2.3\,kB weighs proportionally more.

\section{Conclusions and Future Work}\label{sec:concl}
This paper has proposed a post-quantum migration of the key-establishment mechanisms of OWE and FILS in Wi-Fi networks, replacing the Diffie--Hellman agreement with a hybrid ECDH $+$ ML-KEM-768 scheme, with a KDF-based hybrid combiner ---an instance of a construction proven under the dual-PRF assumption~\cite{bindel2019}--- that remains secure as long as one of the two components survives, and a negotiation design whose downgrade mitigation is specified but, beyond a local policy, not implemented in the prototype. We do not claim an original proof: the contribution is the concrete instantiation over OWE and the verification that the implementation respects the construction (concatenation order, context salt).

The proposal has been validated, \emph{on OWE}, with a reference implementation on \textit{hostapd}/\textit{wpa\_supplicant}, showing that the migration is feasible without adding round trips, fits with ML-KEM-512 and ML-KEM-768 in a single association frame even under a conservative 1500-byte fragmentation threshold (ML-KEM-1024 fragments at that threshold and requires a larger supplicant buffer), and adds no association overhead beyond a few tenths of a millisecond on the testbed and a computational cost (ML-KEM-768) of $\sim$0.19\,ms, marginal on conventional clients and APs. The implementation additionally satisfies the operational preconditions for forward secrecy (per-association ephemeral keys, secure erasure), stated formally with respect to the adversary model. The cryptographic, computational, and forward-secrecy core carries over to FILS, whose frame budget is left for specific measurement (Section~\ref{sec:fils}).

The proposal sits midway between Harkins's protocol-modification approach~\cite{harkins2024} and the application-level overlay of Keh and Chen~\cite{keh2026}. Compared with the overlay, modifying the OWE and FILS key agreement avoids an additional protocol layer, at the cost of requiring changes in the supplicant and the AP.

The main limitations are three. First, the functional validation has been performed on a user-space testbed (\textit{hostapd}/\textit{wpa\_supplicant} with a simulation driver); consequently, physical-layer airtime and throughput are not representative of real radios, and a test on hardware and commercial drivers (Broadcom, Qualcomm, Intel) confirming the in-band fragmentation behavior and measuring the real association latency remains pending. Nonetheless, the frame sizes, the per-threshold fragmentation, and the computational cost are real measurements. Second, the microbenchmark has been measured on a single virtualized platform (a VirtualBox VM with 2 vCPUs) and a specific version of the cryptographic library; absolute values depend on the environment and, in particular, virtualization may add scheduling jitter. Third, the downgrade before the first association remains open and, with the default policy, can even occur without an adversary (Section~\ref{sec:negot}): analogously to the attacks on the WPA3 transition mode, closing it requires a formal specification of which fields are authenticated and when; effective standardization also depends on the progress of P802.11bt~\cite{tgbt}.

Future work includes, as an immediate step, the empirical validation of the frame budget and fragmentation of FILS in its \emph{Shared Key} and \emph{Public Key} variants (Section~\ref{sec:fils}); in addition, porting the prototype to real hardware and drivers to measure in-band association latency and fragmentation, extending the analysis to constrained IoT nodes, formalizing the proof of the combiner in the dual-PRF model, advertising the capability in Beacons, addressing the early-downgrade closure, adding X25519 as the classical component, and tackling the SAE case via a post-quantum PAKE.

\section*{Acknowledgments}
The authors thank the Universidad de Cantabria, MATESCO Department and Eni Plenitude Iberia for their support of this work.

\bibliographystyle{IEEEtran}
\bibliography{references}

\begin{thebibliography}{10}
\providecommand{\url}[1]{#1}
\csname url@samestyle\endcsname
\providecommand{\newblock}{\relax}
\providecommand{\bibinfo}[2]{#2}
\providecommand{\BIBentrySTDinterwordspacing}{\spaceskip=0pt\relax}
\providecommand{\BIBentryALTinterwordstretchfactor}{4}
\providecommand{\BIBentryALTinterwordspacing}{\spaceskip=\fontdimen2\font plus
\BIBentryALTinterwordstretchfactor\fontdimen3\font minus
  \fontdimen4\font\relax}
\providecommand{\BIBforeignlanguage}[2]{{%
\expandafter\ifx\csname l@#1\endcsname\relax
\typeout{** WARNING: IEEEtran.bst: No hyphenation pattern has been}%
\typeout{** loaded for the language `#1'. Using the pattern for}%
\typeout{** the default language instead.}%
\else
\language=\csname l@#1\endcsname
\fi
#2}}
\providecommand{\BIBdecl}{\relax}
\BIBdecl

\bibitem{shor1997}
P.~W. Shor, ``Polynomial-time algorithms for prime factorization and discrete
  logarithms on a quantum computer,'' \emph{SIAM Journal on Computing},
  vol.~26, no.~5, pp. 1484--1509, 1997.

\bibitem{grover1996}
L.~K. Grover, ``A fast quantum mechanical algorithm for database search,'' in
  \emph{Proc. 28th Annu. ACM Symp. Theory of Computing (STOC)}.\hskip 1em plus
  0.5em minus 0.4em\relax ACM, 1996, pp. 212--219.

\bibitem{tan2023}
T.~G. Tan, J.~Zhou, V.~Sharma, and S.~P. Mohanty, ``Post-quantum adversarial
  modeling: A user's perspective,'' \emph{Computer}, vol.~56, no.~8, pp.
  58--67, Aug. 2023.

\bibitem{mosca2018}
M.~Mosca, ``Cybersecurity in an era with quantum computers: Will we be ready?''
  \emph{IEEE Security \& Privacy}, vol.~16, no.~5, pp. 38--41, 2018.

\bibitem{harkins2024}
D.~Harkins, ``Post-quantum cryptography in 802.11,'' IEEE 802.11 Working Group,
  Tech. Rep. IEEE 802.11-24/1103r1, Jul. 2024.

\bibitem{rfc8110}
D.~Harkins and W.~Kumari, ``Opportunistic wireless encryption,'' Internet
  Engineering Task Force, RFC 8110, Mar. 2017.

\bibitem{stebila2024}
D.~Stebila, S.~Fluhrer, and S.~Gueron, ``Hybrid key exchange in {TLS} 1.3,''
  Internet Engineering Task Force, RFC 9954, Jul. 2026.

\bibitem{fils2016}
{IEEE}, ``{IEEE} standard for information technology---part 11: Wireless {LAN}
  {MAC} and {PHY} specifications---amendment 1: Fast initial link setup,''
  IEEE, IEEE Std 802.11ai-2016, 2016.

\bibitem{fips203}
{National Institute of Standards and Technology}, ``Module-lattice-based
  key-encapsulation mechanism standard,'' National Institute of Standards and
  Technology, Gaithersburg, MD, USA, Federal Inf. Process. Stds. (FIPS)
  Publication 203, Aug. 2024.

\bibitem{tgbt}
\BIBentryALTinterwordspacing
{IEEE P802.11bt Task Group}, ``{IEEE} {P802.11bt/D1.0}: Draft
  amendment---enhancements for post-quantum cryptography,'' IEEE 802.11 Working
  Group, WG Technical Letter Ballot 297, Sep. 2026. [Online]. Available:
  \url{https://www.ieee802.org/11/LetterBallots/LB297bt/LB297_email.txt}
\BIBentrySTDinterwordspacing

\bibitem{keh2026}
M.~Keh and Y.~Chen, ``Implementing post-quantum cryptography to industrial
  wireless networks,'' \emph{Electronics}, vol.~15, no.~9, p. 1834, Apr. 2026.

\bibitem{nistcswp39}
{National Institute of Standards and Technology}, ``Considerations for
  achieving crypto agility: Strategies and practices,'' National Institute of
  Standards and Technology, Gaithersburg, MD, USA, NIST Cybersecurity White
  Paper (CSWP)~39, Dec. 2025.

\bibitem{bindel2019}
N.~Bindel, J.~Brendel, M.~Fischlin, B.~Goncalves, and D.~Stebila, ``Hybrid key
  encapsulation mechanisms and authenticated key exchange,'' in
  \emph{Post-Quantum Cryptography (PQCrypto)}, ser. LNCS, vol. 11505.\hskip 1em
  plus 0.5em minus 0.4em\relax Springer, 2019, pp. 206--226.

\bibitem{pqm4}
M.~J. Kannwischer, J.~Rijneveld, P.~Schwabe, and K.~Stoffelen, ``pqm4: Testing
  and benchmarking {NIST} {PQC} on {ARM} {Cortex-M4},'' in \emph{Second NIST
  PQC Standardization Conference}, Santa Barbara, CA, USA, Aug. 2019, current
  ML-KEM cycle counts:
  \url{https://github.com/mupq/pqm4/blob/master/benchmarks.md} (accessed Nov.
  2026).

\end{thebibliography}

\end{document}